\documentclass[sigconf]{acmart}
\usepackage[ruled,vlined]{algorithm2e}
\usepackage{algorithmic}
\usepackage{subfigure}
\usepackage{graphicx}
\usepackage{multirow}
\usepackage{array}
\usepackage{colortbl}
\usepackage{booktabs} 
\usepackage{xcolor}
\usepackage{amsmath}
\usepackage{bm}
\usepackage{amsthm}
\usepackage[most]{tcolorbox}
\usepackage{pifont}

\usepackage[utf8]{inputenc}

\usepackage{cleveref}
\crefname{section}{§}{§§}
\Crefname{section}{§}{§§}
\definecolor{chart Idle}{gray}{.6}
\definecolor{chart Poor}{RGB}{242,28,28}
\definecolor{chart Ok}{RGB}{248,172,37}
\definecolor{chart Ideal}{RGB}{1,151,0}
\definecolor{chart Over}{RGB}{0,125,234}
\definecolor{lightergray}{RGB}{230,230,230}
\definecolor{DarkGreen}{RGB}{30,130,30}

\newdimen\tempdim

\newcommand*{\ChartBox}[3]{%
  \begingroup
    \settoheight{\tempdim}{L}%
    \edef\tempheight{\the\tempdim}%
    \settodepth{\tempdim}{g}%
    \edef\tempdepth{\the\tempdim}%
    \tikz[
      baseline=0pt,
      inner sep=0pt,
    ]
    \node[
      fill={#3!#2},
      rounded corners=1pt,
      anchor=base,
    ]{%
      \vphantom{g\"A}%
      \pgfmathsetlength{\tempdim}{#1}%
      \kern\tempdim\relax
    };%
  \endgroup
}

\AtBeginDocument{%
  \providecommand\BibTeX{{%
    \normalfont B\kern-0.5em{\scshape i\kern-0.25em b}\kern-0.8em\TeX}}}
\AtBeginDocument{%
  \providecommand\BibTeX{{%
    Bib\TeX}}}

\setcopyright{acmlicensed}
\copyrightyear{2018}
\acmYear{2018}
\acmDOI{XXXXXXX.XXXXXXX}

\acmConference[Conference acronym 'XX]{Make sure to enter the correct
  conference title from your rights confirmation emai}{June 03--05,
  2018}{Woodstock, NY}
\acmISBN{978-1-4503-XXXX-X/18/06}


\begin{document}

\title{CreAgent: LLM-enhanced Long-term Evaluation of Recommendation Strategy under Platform Information Asymmetry}

\title{Creator Matters: Evaluating Recommendation Systems under Information Asymmetry}

\title{Enhancing Long-term Recommender Systems Evaluation: A Perspective under Information Asymmetry}

\title{Towards Long-Term Evaluation of Recommender Systems: \\A Perspective under Information Asymmetry}

\title{CreAgent: LLM-Enhanced Creator Simulation for Recommender \\ System Long-term Evaluation Under Information Asymmetry}

\title{CreAgent: Towards Long-Term Evaluation of Recommender \\ System under Platform-Creator Information Asymmetry}


\author{Xiaopeng Ye}
\affiliation{%
\institution{Gaoling School of Artificial Intelligence\\Renmin University of China}
 \city{Beijing}\country{China}
}
\email{xpye@ruc.edu.cn}

\author{Chen Xu}
\affiliation{%
\institution{Gaoling School of Artificial Intelligence\\Renmin University of China}
 \city{Beijing}\country{China}
}
\email{xc_chen@ruc.edu.cn}

\author{Zhongxiang Sun}
\affiliation{%
\institution{Gaoling School of Artificial Intelligence\\Renmin University of China}
 \city{Beijing}\country{China}
}
\email{sunzhongxiang@ruc.edu.cn}

\author{Jun Xu}
\authornote{Corresponding author}
\affiliation{%
\institution{\mbox{Gaoling School of Artificial Intelligence}\\Renmin University of China}
 \city{Beijing}\country{China}
}
\email{junxu@ruc.edu.cn}

\author{Gang Wang}
\affiliation{%
 \institution{Huawei Noah's Ark Lab}
  \city{Shenzhen}
  \country{China}
 }
\email{wanggang110@huawei.com}

\author{Zhenhua Dong}
\affiliation{%
 \institution{Huawei Noah's Ark Lab}
  \city{Shenzhen}
  \country{China}
 }
\email{dongzhenhua@huawei.com}

\author{Ji-Rong Wen}
\affiliation{%
 \institution{\mbox{Gaoling School of Artificial Intelligence}\\Renmin University of China}
  \city{Beijing}
  \country{China}
 }
\email{jrwen@ruc.edu.cn}

\renewcommand{\shortauthors}{Xiaopeng Ye et al.}

%
%
\begin{abstract}




Ensuring the long-term sustainability of recommender systems (RS) emerges as a crucial issue. Traditional offline evaluation methods for RS typically focus on immediate user feedback, such as clicks, but they often neglect the long-term impact of content creators. On real-world content platforms, creators can strategically produce and upload new items based on user feedback and preference trends.
While previous studies have attempted to model creator behavior, they often overlook the role of information asymmetry. This asymmetry arises because creators primarily have access to feedback on the items they produce, while platforms possess data on the entire spectrum of user feedback. Current RS simulators, however, fail to account for this asymmetry, leading to inaccurate long-term evaluations.

To address this gap, we propose CreAgent, a Large Language Model (LLM)-empowered creator simulation agent. By incorporating game theory’s belief mechanism and the fast-and-slow thinking framework, CreAgent effectively simulates creator behavior under conditions of information asymmetry. Additionally, we enhance CreAgent's simulation ability by fine-tuning it using Proximal Policy Optimization (PPO).
Our credibility validation experiments show that CreAgent aligns well with the behaviors between real-world platform and creator, thus improving the reliability of long-term RS evaluations. Moreover, through the simulation of RS involving CreAgents, we can explore how fairness- and diversity-aware RS algorithms contribute to better long-term performance for various stakeholders. CreAgent and the simulation platform are publicly available at \textcolor{blue}{\url{https://github.com/shawnye2000/CreAgent}}.


\end{abstract}


\maketitle


\section{Introduction}

\begin{figure}[t]
    \centering
    \includegraphics[width=0.95\linewidth]{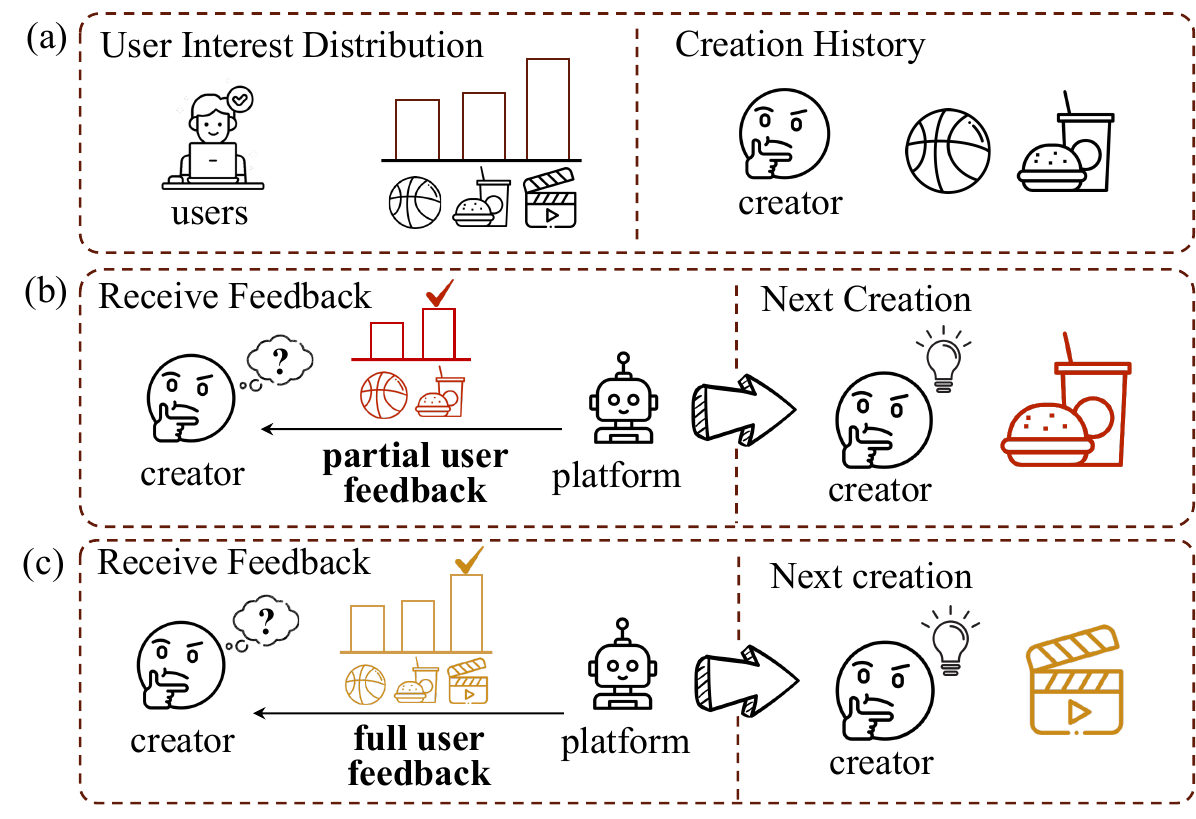}
    \caption{(a) A platform where users favor movies, food, and sports in decreasing order, with a creator who has created sports and food items. A comparison of the creator's creation behavior under (b) limited information; and (c) full information hypothetically.
}
    \label{fig:intro}
\end{figure}

Recently, improving the long-term sustainability of multi-stakeholder recommendation platforms~\cite{abdollahpouri2019multi, burke2016towards_multistakeholder_evaluation}, such as YouTube and TikTok, has emerged as a critical concern~\cite{sun2023take, hohnhold2015focusing, saito2024long_term_off_policy_saito}.
Traditional evaluation methods for recommendation systems (RS) typically focus on metrics that measure the prediction accuracy of immediate user responses (e.g., clicks, purchases)~\cite{zangerle2022evaluating_RS_survey,jarvelin2002ndcg}.
However,  these short-term evaluations often fail to account for long-term effects (e.g., creator retention rate~\cite{el2022quantifying_creator_economy}, long-term user engagement~\cite{zou2019RL4long-term_user_engage}), leading to potential risks for platform developments, such as the emergence of filter bubbles~\cite{nguyen2014explor_filter_bubble}.




\textbf{Creator behaviors under information asymmetry.} When evaluating the long-term impact of RS, it is crucial to consider the content creators~\cite{bhargava2022creator_economy, burke2016towards_multistakeholder_evaluation}.
This is because (1) creators continuously reshape the platform's content ecosystem by uploading items over time; and (2) RS can significantly influence creators' behavior, which in turn impacts the platform's long-term development~\cite{bardhan2022more}. While previous studies have attempted to model the interaction behavior between creators and RS~\cite{ding2024fashionregen, xu2023p, zhang2024large_scale_influencer_selection}, they often overlook the fact that such interaction behaviors are acutally under the condition of \textbf{information asymmetry}~\cite{prasad2023contentprompting, ben2018strategic_content_providers, xu2024ppa}. This asymmetry arises because creators typically have access only to feedback on the items they produced, as restricted by platform policies~\cite{felt2008privacy}, while the platform has access to the full spectrum of feedback data.
To better illustrate this, we will provide an example as follows.



As shown in Figure~\ref{fig:intro} (a), we consider a platform with creators, users, and three genres of items. On such a platform, we assume that the majority of users prefer the genre of movies, followed by a preference for food-related items and then sports-related items.
In a real-world scenario, the platform will only share user feedback on items created by a specific creator directly with that creator. 
Under such partial feedback, the creator that produced sports and food item categories will be more likely to create the new food rather than sports items. This is because the food receives more positive feedback from users (as illustrated in Figure~\ref{fig:intro} (b)). However, if the platform provided him with full user feedback information hypothetically (as illustrated in Figure~\ref{fig:intro} (c)), the creator would likely opt to create a movie item instead. Since the movie is welcomed by most users in the platforms.
Therefore, modeling the information asymmetry is important for simulating the creator's strategic behaviors. It is also highlighted by existing mechanism design studies~\cite{ben2018strategic_content_providers, prasad2023contentprompting}.

\textbf{Simulator for RS long-term evaluation.} Although online test~\cite{kohavi2015onlineabtest} can effectively assess the performance of RS in environments with creators, its high costs hinder the evaluation efficiency~\cite{zangerle2022evaluating_RS_survey}.
Recognizing the need for cost-efficient solutions, recent studies~\cite{zhang2024agent4rec,wang2023recagent, ie2019recsim} have identified recommendation simulators as an ideal approach for affordable long-term evaluation. However, existing simulators~\cite{zhang2024agent4rec,shi2019virtualtaobao, ie2019recsim} often focus exclusively on user behavior simulation, neglecting the impact of creators.

To achieve realistic creator behavior simulations under platform-creator information asymmetry scenarios, we propose a Large Language Model (LLM)-empowered creator agent, called \textbf{CreAgent}. Inspired by the game theory~\cite{prasad2023contentprompting}, we employ a belief module to reflect the limited information status of creators and utilize the fast-and-slow thinking mechanism~\cite{kahneman2011thinkingfastandslow} to capture the thought process under such information asymmetry. Moreover, to better help CreAgent understand the limited user feedback information, we utilize Proximal Policy Optimization (PPO)~\cite{schulman2017ppo} to fine-tune the CreAgent. 
Credibility validation experiments have confirmed CreAgent's ability to exhibit human-like strategic creator behaviors, supported by insights from real-world data patterns and behavior economics principles~\cite{conlisk1996boundedrational, kahneman2013prospecttheory}.

Equipped with the CreAgent, we design a simulated platform environment, which contains highly extensible recommendation models and modified widely-used user agents RecAgent~\cite{wang2023recagent}. 
The simulated platform environment is built using a large-scale dataset that we collected through web crawling from the YouTube website. 
Based on such a dataset, our experiments confirm that our proposed CreAgent can well align with the behaviors of real content creators under information asymmetry interaction patterns. 
Furthermore, utilizing this simulation environment, we reassess various fairness- and diversity-aware RS algorithms to analyze their long-term performance for users, creators, and the platform separately.

\textbf{Contributions.} Our contribution can be summarized as follows:

(1) We emphasize the significance of considering creator strategic behavior under platform-creator information asymmetry scenario during the long-term evaluation of RS.  

(2) We introduce CreAgent, an LLM-powered agent that simulates real-world creator interactions, and an extensible simulation platform for evaluating long-term RS performance with diverse models and user agents.

(3) Extensive experiments demonstrated that our simulation platform effectively replicates real-world RS, enabling a thorough reassessment of the long-term performance of various RS algorithms.

\section{Related Work}

\textbf{Evaluation of RS}. How to evaluate RS is a complex and essential task, which can be divided into short-term and long-term evaluation based on their objectives~\cite{zangerle2022evaluating_RS_survey}. Most current studies focus on the short-term objective using offline metrics~\cite{jannach2020escaping, jannach2012recommender_landscape}, relying on pre-collected log data containing users' explicit (e.g., ratings) or implicit (e.g., click) feedback to compute metrics like prediction accuracy~\cite{willmott2005MAE} and ranking metrics~\cite{jarvelin2002ndcg}. 
Existing fairness-aware~\cite{ye2024bankfair, xu2024taxation, naghiaei2022cpfair}, debiasing~\cite{chen2023bias}, and diversity-aware~\cite{carbonell1998mmr} recommendation research often evaluates using domain-specific indicators based on these short-term metrics (e.g., fairness metric~\cite{gastwirth1972giniindex}, ), which fail to adequately reflect their long-term benefits and how they influence RS in the long run, particularly in multi-stakeholder platform environments~\cite{surer2018multistakeholder}.
Due to the inefficiency and high cost of long-term online A/B test~\cite{kohavi2015onlineabtest, saito2024long_term_off_policy_saito}, offline long-term evaluation has gained significant attention in recent years, which can be divided into two main categories: (1) use short-term metric~\cite{hohnhold2015focusing} or data~\cite{saito2024long_term_off_policy_saito} to predict long-term performance, (2) create an RS simulator to replicate the real-world environment~\cite{ie2019recsim, zhang2024agent4rec, wang2023recagent} for evaluation.
In this paper, we focus on the second type. 

\textbf{RS simulator for long-term evaluation.} 
Most existing recommendation simulators (e.g., LLM-based simulator~\cite{wang2023recagent, zhang2024agent4rec}, reinforcement learning (RL)-based simulator~\cite{ shi2019virtualtaobao,ie2019recsim}) focus on user simulation while overlooking creators, making it difficult to capture the long-term dynamic of content platforms.
Some data-driven methods are proposed to conduct creator simulation.  SimuLine~\cite{zhang2023simuLine} applied heuristic methods to determine creators' next creation in news recommendation. 
Some works~\cite{mladenov2021recsimng, mladenov2020optimizing_long-term} assumed that creators will leave the platform if their user engagement falls below a certain threshold.
Other modeling methods used user positive feedback (e.g., click) as the gradient to update the creation state~\cite{lin2024cfd, yao2024uwo, zhan2021toward_content_provider_rec}.
However, these approaches failed to align with real creation behavior because: (1) they are unable to produce authentic content (e.g., text), instead relying on embeddings to represent the content they create~\cite{zhang2023simuLine}; (2) they cannot capture the personalization of real-world creators; (3) they ignored human behavior patterns under information asymmetry, such as risk aversion in prospect theory~\cite{kahneman2013prospecttheory} and bounded rationality~\cite{selten1990bounded_rationality}.

\textbf{Behaviors under information asymmetry.} Creator behaviors in information asymmetry conditions
have been actively studied and emphasized in the game theory literature~\cite{yao2024uwo, lin2024cfd, xu2024ppa}. 
They typically assume that creators are totally rational, i.e., aiming to maximize their utility, which often lacks personalization and differs from real-world human behavior under risk (i.e., bounded rational~\cite{selten1990bounded_rationality}). Rule-based~\cite{xu2024ppa} and heuristic method~\cite{yao2024uwo, lin2024cfd} are applied to model the strategic behavior. These studies mainly focus on the competition among creators~\cite{yao2024uwo,lin2024cfd} and the design of better platform mechanisms\cite{mladenov2020optimizing_long-term,prasad2023contentprompting} to maximize user welfare.



\section{PRELIMINARIES}
\label{sec:preliminaries}


\textbf{Simulation platform.} In RS, we use $\mathcal{P} =\{\mathcal{U}, \mathcal{I},  \mathcal{C}, \mathcal{G}, \mathcal{D}, \mathbf{E}, 
\mathbf{Y}\}$ to denote the information known to the platform. $\mathcal{U} = \{u\}$, $\mathcal{I} = \{i\}$, $\mathcal{C}=\{c\}$, $\mathcal{G}=\{g\}$ is the set of users, items, creators, and item genres. Each item $i \in \mathcal{I}_g$ in RS belongs to a genre $g\in \mathcal{G}$. An item $i$ created by creator $c$ denotes as $i\in \mathcal{I}_c$.  $\mathbf{E}$, $\mathbf{Y}$ is the user-item exposure matrix and interaction matrix.

At each time step $n \in [1,2,\cdots, N]$, some users $\mathcal{U}^n$ will visit the RS, and some creators $\mathcal{C}^n$ will create items. Each created item $i$ will be added to the item pool $\mathcal{I}$ of RS. The items created in time step $n$ is denoted as $i\in \mathcal{I}^n$.
 For each user $u$, the RS generates a list of $K$ recommended items chosen from $\mathcal{I}$. Each user $u$ can choose to click or skip these items. Then, the RS records the click and exposure interaction in $\mathbf{Y}(n)$ and $\mathbf{E}(n)$, which denotes the exposure matrix and interaction matrix of time step $n$. Specifically, $\mathbf{Y}_{u, i}(n)=1$ if user $u$ clicks item $i$ at time step $n$ and 0 otherwise, $\mathbf{E}_{u, i}(n)=1$ if item $i$ is exposed to user $u$ at time step $n$ and 0 otherwise. Also, RS will save the interaction record $\left( u, i, \mathbf{Y}_{u,i}(n)\right)$ into the interaction record $\mathcal{D}$. For every $T$ step, the RS will be re-trained using interactions in $\mathcal{D}$. 

At the beginning of our simulation, we utilize the dataset $\mathcal{P}^0 =\{\mathcal{U}^0, \mathcal{I}^0, \mathcal{C}^0, \mathcal{G}^0, \mathcal{D}^0\}$ collected from real-world RS to initialize the simulation environment. Specifically, the user's profile, creator's profile and initial belief will be initialized and the RS will be trained using the real-world interactions $\mathcal{D}^0$. During simulation, we use the first $n\in[1,2,\cdots, N_0-1]$ time steps to initialize a stable platform environment that reflects the real-world RS.
Then, the subsequent $n\in[N_0, \cdots, N]$ steps are used to evaluate different RS models.

\textbf{Information asymmetry.} In real-world platforms, such as YouTube, the platform can access comprehensive user behavior data (e.g., watch time, click/like history), policies~\cite{stein2013policy_youtube} and incentives~\cite{prasad2023contentprompting}, which is not disclosed to creators. Meanwhile, creators can only infer whether their content aligns with user preferences by analyzing feedback (e.g., comments, likes, and other interactions) on their created items and then adjust their creation strategies accordingly. Thus, significant information asymmetry exists in the platforms, as highlighted by previous studies~\cite{lin2024cfd, yao2024uwo, prasad2023contentprompting}. 

In this paper, we denote the platform-possessed information as $\mathcal{P}$, whereas each creator $c$ has access to a subset of this information, denoted as $\mathcal{P}_c= \{\mathcal{I}_c,  \mathbf{E}^\top_{i} ,  \mathbf{Y}^\top_{i} \mid i\in \mathcal{I}_c\}, \forall c\in \mathcal{C}$. Since $\mathcal{P}_c \neq \mathcal{P}$, this disparity indicates that there is an information asymmetry between the platform and the creators.

\textbf{Organization.} To evaluate the long-term impact of RS under the platform-creator information asymmetry environment, we propose a simulation platform with creator behavior modeling. The overview of the simulation platform is illustrated in Figure~\ref{fig:overview}. The simulation platform can be decoupled into the LLM-empowered creator agent \textbf{CreAgent} (see Section~\ref{sec:creagent} and Figure~\ref{fig:overview}~(a)) and the \textbf{platform environment} (see Section~\ref{sec:platform} and Figure~\ref{fig:overview}~(b)). Afterwards, we conduct experiments to demonstrate the credibility of our proposed CreAgent (see Section~\ref{sec:creator_distibution_alignment}). Finally, we use CreAgent and the simulation platform to assess the long-term impact of RS models on diverse stakeholders (see Section~\ref{sec:RS_evaluation}).

\section{CreAgent}
\label{sec:creagent}

\begin{figure*}
    \centering
    \includegraphics[width=0.95\linewidth]{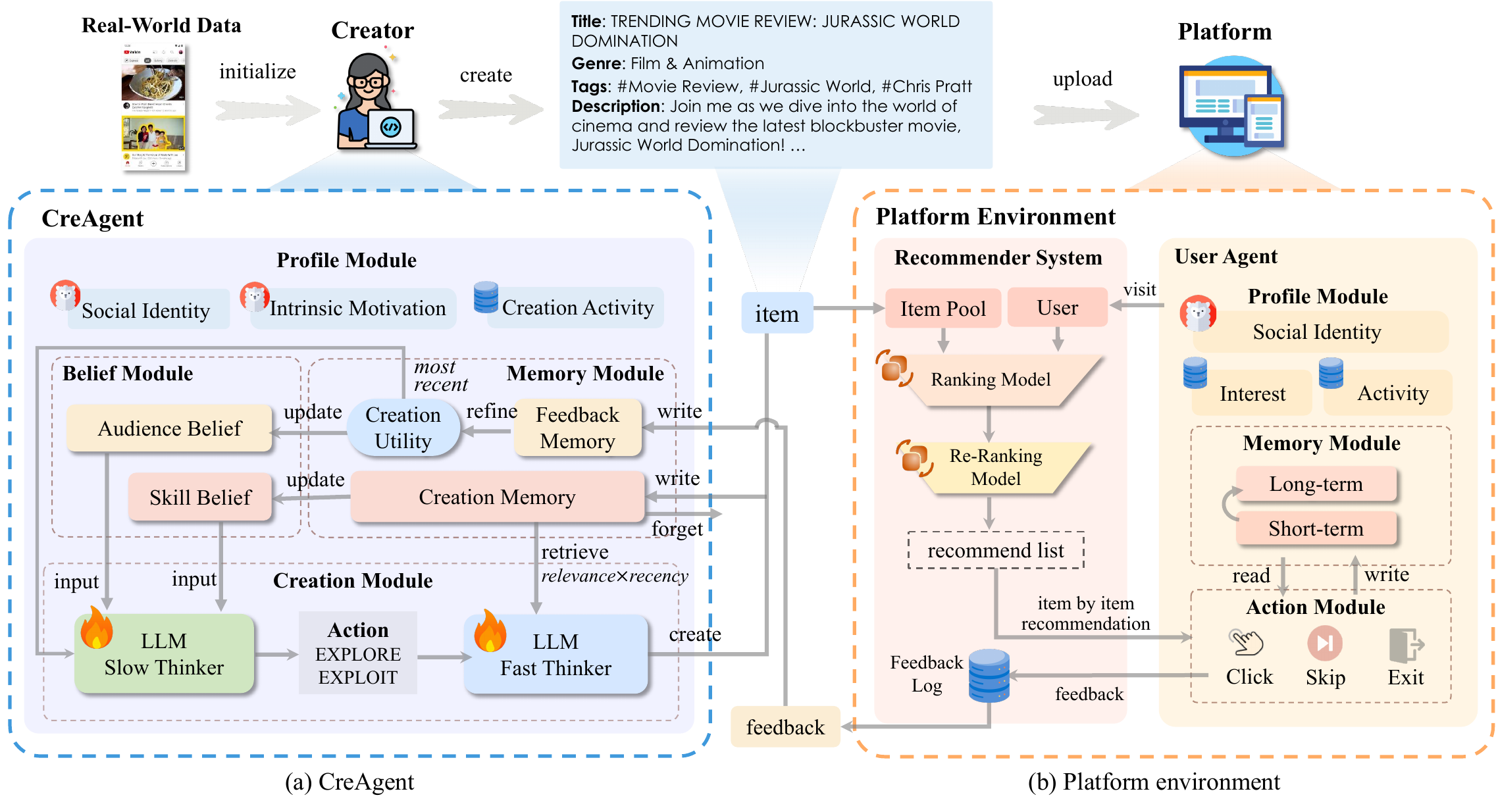}
    \caption{The overall workflow of our simulation platform.
 (a) CreAgent, initialized with the real-world YouTube dataset, employs a belief mechanism combined with fast-and-slow thinking to create the next item strategically based on platform-provided user feedback.
  (b) Platform environment, which consists of an extensible two-stage recommendation system and modified widely-used user agent, collects item feedback from users and sends it to CreAgent. }
    \label{fig:overview}
\end{figure*}

In this section, we leverage the human-like analyzing, generation capabilities, and world knowledge~\cite{tamkin2021understanding_capbility_of_LLM} of LLMs to simulate real-world strategic behavior of creators under information asymmetry scenarios. As demonstrated in Figure~\ref{fig:overview} (a), CreAgent consists of four modules: the profile and memory module, the game-theory-inspired belief module, and the creation module integrating fast-and-slow thinking~\cite{kahneman2011thinkingfastandslow}. Finally, we utilize PPO-based fine-tuning to enhance CreAgent's analytical and creative capabilities under limited information.

\subsection{Profile Module Design}
In this module, we initialize CreAgent's profile using a pre-collected real-world dataset, crucial for aligning agent behavior with real human actions. CreAgent's profile comprises three elements: social identity, intrinsic motivation, and creation activity.

\textbf{Social identity.} In reality, content creation, as social behavior, is driven by not only economic interests but also social identity, a factor often overlooked in behavior modeling~\cite{yao2023howbad, yao2024userwelfare_opt, zhan2021toward_content_provider_rec}. To address this, we employ an LLM to identify creators' social identities by analyzing their creation history $\mathcal{I}^0_c$ and basic information. 
For example, a creator who enjoys producing lifestyle content might be identified as a ‘‘lifestyle advocate'', while one who frequently creates gaming content could be summarized as a ‘‘gaming enthusiast''.
We represent this social identity as $P^s_c$, whose format is in text, and the detailed format can be found in Appendix~\ref{app:prompt}.

\textbf{Intrinsic motivation.}
Established studies~\cite{bi2020intrinsic_moti, hull1943drive_theory} show that intrinsic motivation is a key factor that affects creators' creation behavior, yet this motivation varies from person to person. For example, while some creators frequently produce content to obtain more revenue, others create simply to share their lives without high-profit expectations. We use LLM to help us summarize each creator's intrinsic motivation $P^m_c$ from their creation history $\mathcal{I}^0_{c}$ and frequency as one part of their profiles.
The format $P^m_c$ of is in text, and the detailed format can be seen in Appendix~\ref{app:prompt}.

\textbf{Creation activity.}
In addition to the above two economic characteristics, creation activity (i.e., the frequency of a creator’s creation) is also an important trait of the creator. We use the collected dataset $\mathcal{I}_c^0$ to initialize the inherent activity level $p^a_{c}$ for each creator $c$. It represents the average number of items created by each creator every day.
After getting $p^a_{c}$ of each creator, the creation probability for each creator at each step is $p^a_{c}/\eta$ where $\eta = \max\{p^a\}$ is the maximum activity among all creators.

\subsection{Memory Module Design}
 Considering the two key pieces of information that creators prioritize and store in reality, we build two memories for CreAgent: feedback memory and creation memory.

\textbf{Feedback memory.}
Different from user agent memory modeling~\cite{wang2023recagent}, the user feedback received by each creator is usually recorded and maintained by the platform over a considerable period~\cite{lange2007publicly_private}. We denote the feedback memory of creator $c$ as $\mathcal{M}_c^{\text{feed}}$. Due to its position in a state of information asymmetry,
at the end of each time step $n$, the feedback memory 
$\mathcal{M}_c^{\text{feed}}$ will be updated based on the partial user feedback information provided by the platform on history creation:
\begin{equation}
  \mathcal{M}_c^{\text{feed}} = \mathcal{M}_c^{\text{feed}} \cup \left\{ \sum_{u \in\mathcal{U}^n} \mathbf{E}_{u,i}(n), \sum_{u \in\mathcal{U}^n} \mathbf{Y}_{u,i}(n) \mid i \in \mathcal{I}_c \right\}   
\end{equation}

After storing user feedback in memory $\mathcal{M}_c^{\text{feed}}$ at time step $n$, the feedback memory undergoes a refinement process to calculate the utility of each created item. Specifically, it refers to the time-average exposure and clicks on each item $i\in \mathcal{I}_c$ up to the present step $n$.
{\small
\begin{equation}
    \mathbf{z}_i(n) = \frac{1}{\theta}  \left[\beta \sum_{k=t(i)}^{n} \sum_{u \in\mathcal{U}^k} \mathbf{E}_{u,i}(k)   + (1- \beta) \sum_{k=t(i)}^{n} \sum_{u \in\mathcal{U}^k} \mathbf{Y}_{u,i}(k) \right] ,
\end{equation}}
where $\theta = \frac{1}{n-t(i)+1}$ and $t(i)$ is the creation time step of item $i\in \mathcal{I}_c$, $\beta$ is the hyper-parameters that control the importance of exposure for the creator. This utility can be configured by adjusting $\beta$ according to specific scenarios, either as exposure-based~\cite{xu2023p, patro2020fairrec} or click-based~\cite{yao2024uwo}. 
The utility $\mathbf{z}$ will be used as input for both the belief module (Section~\ref{sec:belief_module}) and the creation module (Section~\ref{sec:creation_module}) to help CreAgent make decisions.

\textbf{Creation memory.}
The creation memory $\mathcal{M}_c^{\text{cre}}$ stores the historical creation item information of the creator agent. To better reflect the human time-fading memory mechanism, we draw on prior studies~\cite{wang2023recagent} and incorporate a power function forgetting rate~\cite{wang2023recagent}.
Before each creation, creators will retrieve the most relevant and recent creation experience from the creation memory. After creation at time step $n$, the newly created item will be saved to the creation memory:
$
\mathcal{M}_c^{\text{cre}} 
 = \mathcal{M}_c^{\text{cre}} \cup \{i \mid i \in \mathcal{I}_c\cap \mathcal{I}^n\},
$ which is used to update the skill belief (Section~\ref{sec:skill_belief}) and as the input of fast thinker~(Section~\ref{sec:fast_thinker}) for future creation.

\subsection{Belief Module Design}
\label{sec:belief_module}
Under information asymmetry conditions, creators usually do not have exact information about other stakeholders (e.g., user preferences, and item genres), and their creation behavior is often driven by their limited information expectations about it (i.e., beliefs).
In this module, we draw inspiration from game theory~\cite{prasad2023contentprompting} and employ the belief mechanism to reflect the limited information status as a guidance for the strategic creation of CreAgent.
 The beliefs of CreAgent are typically formed based on previous information provided by the platform and are updated over time and with the acquisition of new information.
Specifically, the strategic creation behavior of CreAgent is mainly driven by two types of beliefs: skill belief and audience belief.

\textbf{Skill belief.} 
\label{sec:skill_belief}
Under limited information, creators have incomplete knowledge of item genres. They will gradually acquire more genre information and improve their creation proficiency in those genres during the creative process.
The skill belief represents the creator’s confidence in their ability to create each genre of items, which is defined as the percentage of their content created in each genre. 
At the start of each time step $n$, the skill belief of creator $c$ for genre $g$ will be updated based on the creation memory $\mathcal{M}^{\text{cre}}_c$: 
\begin{equation}
\mathbf{B}^{\text{skill}}_{c,g}(n) =  \frac{|\mathcal{I}_{g} \cap \mathcal{I}_c|}{|\mathcal{I}_{c}|}, \quad \forall g \in \mathcal{G}, \mathcal{I}_{g},\mathcal{I}_{c} \subset \mathcal{M}^{\text{cre}}_c.
\end{equation}
Due to some creators being more adept at producing certain types of content than others (e.g., due to specific talent, interests, and facilities), we initialize the skill belief $\mathbf{B}^{\text{skill}}_{c}(1)$ using the pre-collected dataset $\mathcal{I}_c^0$.

\textbf{Audience belief.} The audience belief represents the creator's internal understanding and expectation of user preferences in each genre. At the start of each time step $n$, the audience belief of creator $c$ in genre $g$ will be updated based on the user feedback stored in the feedback memory $\mathcal{M}^{\text{feed}}_{c}$:
\begin{equation}
\mathbf{B}^{\text{aud}}_{c,g}(n) = \frac{\sum_{i\in \mathcal{I}_{g} \cap \mathcal{I}_c} \mathbf{z}_i(n)}{|\mathcal{I}_{g} \cap \mathcal{I}_c|},\quad \forall i \in \mathcal{I}_c.
\end{equation}
Similarly to the skill belief, we initialize the audience belief $\mathbf{B}^{\text{aud}}_{c}(1)$ using the real-world creator's history items $\mathcal{I}_c^0$.


\subsection{Creation Module Design}
\label{sec:creation_module}
In this module, we apply the fast-and-slow thinking mechanism~\cite{kahneman2011thinkingfastandslow, yao2024hdflow, lin2024swiftsage} to equip our creator agents with human-like analysis and creation abilities. The thought process is composed of two phases~\cite{kahneman2011thinkingfastandslow}: slow thinking for strategic planning and analysis, and fast thinking for rapid content generation based on experience and instinct.

\textbf{Slow thinker.}
During each creation,  user feedback on the most recently created item directly affects the creator's judgment on whether to continue with the current creative strategy or change it.
At the start of each time step $n$, creator $c\in \mathcal{C}^n$ gets three key factors that affect his creation strategy as inputs: (1) the utility of the most recently created item $i$, i.e.,$\mathbf{z}_i(n)$
, (2) the skill belief $\mathbf{B}^{\text{skill}}_{c}$ and audience belief $\mathbf{B}^{\text{aud}}_{c}$, and (3) the social identity $P^s_c$ and intrinsic motivation $P^m_c$. Then, these inputs are then fed into the LLM via the designed Chain-of-Thought~\cite{wei2022cot} prompt $P_1$ for slow thinking:
\begin{equation}
\mathbf{A}^{\text{exp}}_c(n) = \text{LLM}\left[ P_1\left(g(i), \mathbf{z}_i(n)
,\mathbf{B}^{\text{aud}}_{c}(n), \mathbf{B}^{\text{skill}}_{c}(n), P^s_c, P^m_c \right) \right],
\end{equation}
where  $i = h(\mathcal{I}_c)$ represent the most recent item created by creator $c$, where $h(\cdot)$ fetches the new item within $(\cdot)$, and $g(i)$ denotes the genre of item $i$. $\mathbf{A}^{\text{exp}}_c(n)$ is the text-based explore/exploit action taken by the creator agent $c$ at time step $t$. Specifically, CreAgent will choose whether to continue creating within the current genre $g(i)$ or to explore another genre $\mathcal{I}/\{g(i)\}$. The detailed format of prompt $P_1$ and action $\mathbf{A}^{\text{exp}}_c$ is in Appendix~\ref{app:prompt}.

\textbf{Fast thinker.} 
\label{sec:fast_thinker}
After generating analytical results, the creator agent will produce content based on these findings. The content is primarily divided into four sections: item title, item genre, item tags, and item description. Before creating content, the creator agent will retrieve the creation experience $f(\mathcal{M}^{\text{cre}}_c)$ from the creation memory $\mathcal{M}^{\text{cre}}_c$ based on the action $\mathbf{A}^{\text{exp}}_c(n)$, to assist the fast thinker in the creation process.
\begin{equation}
    \mathbf{A}^{\text{cont}}_c(n) = \text{LLM}\left[ P_2\left(f\left(\mathcal{M}^{\text{cre}}_c, \mathbf{A}^{\text{exp}}_c(n)\right), P^s_c, P^m_c\right)\right],
\end{equation}
where $P_2$ is the designed prompt for fast thinking. The item $\mathbf{A}^{\text{cont}}_c(t)$ is the content of the newly created item, which contains time title, genre, tags, and description, $f(\cdot_1, \cdot_2)$ is the retriever which retrieves the most relevant and recent creation history from $(\cdot_1)$ based on $(\cdot_2)$. The detailed format of $P_2$ and $\mathbf{A}^{\text{cont}}_c$ is in Appendix~\ref{app:prompt}.

\subsection{Fine-tuning to Enhance Creative Ability}
\label{sec:ppo}
In this section, we use the Proximal Policy Optimization (PPO) algorithm~\cite{schulman2017ppo} to fine-tune CreAgent.  By simulating the real-world creation cycle of creating, receiving rewards, analyzing, and creating again, we aim to:
(1) improve CreAgent's understanding of creators' limited information status and user feedback; (2) enhance CreAgent's analytical and creative capabilities.

\textbf{Reward modeling.}
In real platform scenarios, creators receive user feedback on their created items from the platform and benefit from it, which also helps them adjust their existing creation strategies. In this paper, we utilize the PPO algorithm and treat the the platform environment (Section~\ref{sec:platform}) as the reward model, to replicate the process through which creators learn their creative strategies in the real world. For each created item $i\in\mathcal{I}_c$, the reward is the weighted utility until the current time step $n$, i.e., $\lambda \mathbf{z}_i(n)$,
where $\lambda$ is a constant coefficient used to ensure training stability by preventing the reward from being too large or too small.

\textbf{Replay buffer.}
In real-world scenarios, user feedback is often not collected immediately but takes some time to accumulate. After the creation of item $i$ at time step $n=t(i)$, we first store the state $s^c_n=(\mathbf{B}^{\text{skill}}_{c,g}(n), \mathbf{B}^{\text{aud}}_{c,g}(n))$) and action $a^c_n=(\mathbf{A}^{\text{exp}}_c(n), \mathbf{A}^{\text{cont}}_c(n))$ in the log. After accumulating $N_r$ steps util step $n+N_r$, we save the reward $r^c_{n}= \lambda \mathbf{z}_i(n+N_r)$, state, and action related to the item $(s^c_{n}, a^c_{n}, r^c_{n})$ into the replay buffer. For every $N_u$ step, we update the policy by sampling $M$ records from the replay buffer.


\textbf{PPO optimization.}
To avoid the LLM policy $\pi_{\mathbf{\theta}}$ with parameter $\theta$ deviating from the initial reference policy $\pi_{\mathbf{\theta}_{\text {init}}}$ too far, we follow~\cite{ouyang2022training} and introduce the KL-divergence penalty into the current reward function. Therefore, the policy optimization formula is:
\begin{equation}
\mathcal{L}_{P P O}=\mathbb{E}\left[\lambda \mathbf{z}_i(n)-\beta K L\left(\pi_{\mathbf{\theta}_{\text {init}}}, \pi_{\mathbf{\theta}}\right)\right],
\end{equation}
where $\lambda$ is the hyper-parameter to prevent final reward $\lambda \mathbf{z}_i(n)$ too high to train stably.


\section{Platform Environment}
\label{sec:platform}
In this section, we introduce a simulated platform environment that mirrors the one creators engage with. This environment is primarily for receiving items uploaded by creators and providing them with user feedback. As illustrated in Figure~\ref{fig:overview} (b), the platform environment is mainly divided into the recommendation system and user agents.

\textbf{Recommender system modeling.}
The RS of the simulation platform primarily focuses on extensibility. As shown in Figure~\ref{fig:overview}, it encompasses a dynamic item set that allows items to be added freely, a user set, a two-stage recommendation process that includes ranking and re-ranking, and a feedback log to store user feedback.
We discuss two aspects of environment construction that resonate with real-world RS, including two-stage ranking, and item-by-item recommendation.

\textbf{Two-stage ranking.}
In real industrial scenarios, recommendation lists are often generated through a multi-stage ranking process. We also consider this situation by allowing both the ranking and re-ranking algorithms to be flexibly replaced according to specific requirements. Through this design, the platform can flexibly adjust and evaluate different recommendation strategies, whether by applying only the ranking model (e.g., BPR~\cite{rendle2012bpr}) or by considering long-term objectives (e.g., fairness~\cite{wang2023surveyonfairness} and diversity~\cite{kunaver2017diversity_survey}) in the re-ranking stage.

\textbf{Item-by-item recommendation.} After generating the recommendation list, RS will recommend items to users on an item-by-item basis. This scenario is currently widely used in online recommendation platforms such as YouTube~\cite{song2012youtube} and TikTok.

\textbf{User agents.}
Given the maturity of existing user simulator~\cite{wang2023recagent, zhang2024agent4rec, shi2019virtualtaobao}, we utilize the widely-used LLM-based user simulator RecAgent~\cite{wang2023recagent} as our user agent. RecAgent effectively simulates the human memory mechanism and short- and long-term interests by integrating sensory, short-term, and long-term memory, enabling it to simulate user behavior over time.
Also, to better reflect users' long-term behavior, we make some targeted changes. 
For the profile, we define a targeted social identity for each user, summarized by the LLM based on their interaction and comment history. The user's long-term interests and activity levels are also derived from pre-collected real-world dataset $\mathcal{U}^0$.
After being recommended, the user agent can take three actions: click the item, skip the item, or exit the platform. To ensure the credibility of the platform environment, we also conduct experiments to verify how effectively the interaction behavior of user agents aligns with that of real users (please see Appendix~).


\section{Creator Credibility Evaluation}
\label{sec:creator_distibution_alignment}
In this section, we conduct experiments to verify the effectiveness of CreAgent in long-term simulation by addressing the following research questions:
\\
\textbf{RQ1}: Can CreAgent well align with the real-world creator patterns?
\\
\textbf{RQ2}: Can our simulation platform well simulate the information asymmetry conditions in the real world?
\\
\textbf{RQ3}: What is the computational cost of our simulation platform?



\subsection{Experimental Setups}
\textbf{Real-world dataset collection.}
We initiate a focused data collection effort to tackle the lack of recommendation datasets that include detailed information about content creators and their items. We leverage the YouTube Data API to collect item information and comment data from channels in \textit{YouTube}~\cite{song2012youtube}, the world's leading content platform, known for its diverse and influential content creators. This choice is strategic to ensure our simulation closely mirrors the real-world platform.
Specific details of the dataset are provided in Appendix~\ref{sec:dataset}.

\textbf{Simulation setups.} 
We utilize two A6000 GPUs for simulations, setting the number of user agents $|\mathcal{U}|=100$ and creator agents $|\mathcal{C}|=50$, with a total of $N=100$ simulation steps. All agents are powered by the Llama3-8B~\cite{touvron2023llama}, updated and executed in parallel using multi-threading. The recommendation list length $K=5$.  The RS model is trained on the real-world dataset $\mathcal{D}^0$ and retrained every $T=5$ step using all interaction data $ \mathcal{D}^0\cup\mathcal{D}$ to reflect the periodic update of real-world RS. 
This section uses the Deep Interest Network (DIN)~\cite{zhou2018din} as the base RS model for the credibility evaluation.
To better align with real-world content platform scenarios~\cite{zhang2017timeliness, liang2024ensure_timeliness} and help to cold-start new items, we consider the timeliness of recommendation by removing excessively outdated items (older than $\eta=20$ steps) from the item recommendation pool.

\textbf{Baselines.} We select several classic creator behavior simulation baselines, where the baseline methods determine the genre of the next created item during each time step. Creator Feature Dynamics (\textbf{CFD})~\cite{lin2024cfd, zhan2021toward_content_provider_rec}: creators adjust their creation strategy using user feedback as the gradient, scaled by a learning rate.
Local Better Response (\textbf{LBR})~\cite{yao2024uwo}: creators generate a random direction and evaluate its utility (impact on user feedback). If beneficial, they update their strategy incrementally; otherwise, they maintain the current strategy. 
\textbf{SimuLine}~\cite{zhang2023simuLine}: creators' next creation is determined through probabilistic sampling based on the number of likes from previous steps. 

\subsection{\mbox{RQ1: Real-world Dataset Alignment}}
\label{sec:dataset_alignment}
We first evaluate whether the data pattern simulated by CreAgent aligns with the real-world creator pattern, which ranges from four key aspects: preferences, diversity, activity, and content. Then, we conduct an ablation study to reveal the importance of key modules.




\begin{figure}
    \centering
    \includegraphics[width=1\linewidth]{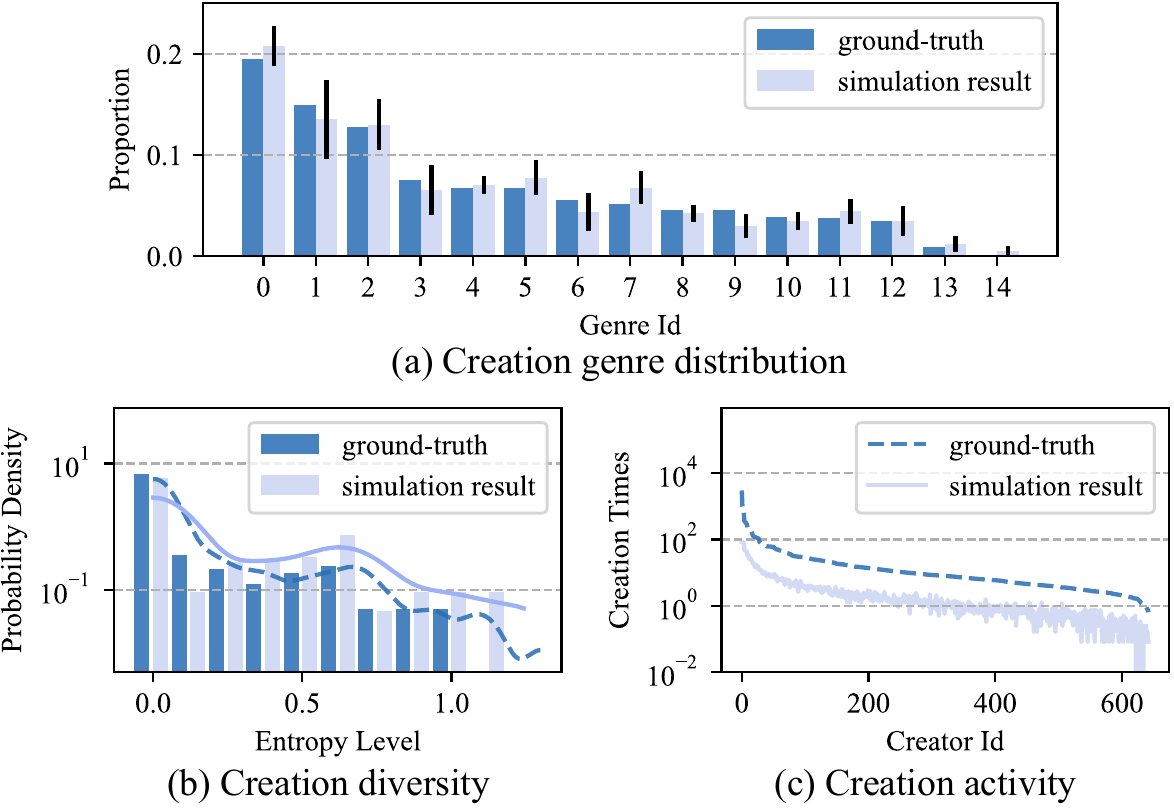}
    \caption{Comparison between the creation genre preference, diversity, and activity of the ground-truth and agent-simulated result.}
    \label{fig:creation_align}
\end{figure}

\begin{table}[!ht]
    \centering
    \caption{Comparison of the divergence between the simulated and real-world distributions using Jensen-Shannon divergence~\cite{menendez1997JS_div}, with genre-level for preference evaluation and individual-level for diversity.}
    \begin{tabular}{l|cc}
    \toprule
      Creator Modeling Method &Preference & Diversity \\ \midrule
      Creator Feature Dynamics~\cite{lin2024cfd, zhan2021toward_content_provider_rec} & 0.2537 & 0.7204 \\ 
       Local Better Response~\cite{yao2024uwo} 
       & 0.2833 & 0.6284 \\ 
SimuLine~\cite{zhang2023simuLine} 
& 0.3175 &  0.6949\\ \hline
       CreAgent (w/o Audience Belief) & 0.2671 & 0.6159 \\ 
        CreAgent (w/o Skill Belief) & 0.2928 & 0.5532 \\ 
        CreAgent (w/o Fast-Slow Thinker) & 0.1728 & 0.5638 \\ 
       \textbf{CreAgent} & $\mathbf{0.1667}$ & $\mathbf{0.3014}$ \\ \bottomrule
    \end{tabular}
    \label{tab:baseline}
\end{table}

\subsubsection{\textbf{Creation Preference Alignment}}
In real platforms, each creator has diverse preferences that influence their content and shape the platform’s ecosystem. To accurately replicate this in our simulation, we first need to
ensure that the simulated creators’ preference distribution aligns with the real-world dataset

As shown in Figure~\ref{fig:creation_align} (a), we plot the creation genre distribution of the YouTube dataset (i.e., the light blue bar) and our simulation (i.e., the deep blue bar). Specifically, the x-axis represents item genres, and the y-axis is the proportion of creation times all CreAgents created in each genre during the first 10 environment initialization steps.
From the comparison, we observe that our creator agents ultimately achieved a genre distribution comparable to the real-world dataset, effectively replicating a similar content ecosystem. 
Additionally, we notice some differences in certain categories (e.g., CreAgents show a stronger preference for genre 7 and less for genre 6 compared to real creators), which we attribute to the influence of the LLM’s pre-trained knowledge. Table~\ref{tab:baseline} demonstrates that CreAgent exhibits greater consistency with the dataset’s preference distribution compared to the baselines.

\subsubsection{\textbf{Creation Diversity Alignment}}

Creator's creation diversity is also a key aspect. Some real-world creators focus on a single category of items, while others actively explore different categories. 
Figure~\ref{fig:creation_align} (b) presents a histogram of the creation diversity for creators on the simulated platform and those in the YouTube dataset. We use the entropy of genre frequencies to represent each creator’s diversity (the x-axis). The solid lines in the figure represent the Kernel density estimation (KDE) curves~\cite{wkeglarczyk2018kde}. We observe that CreAgent effectively replicate the diversity distribution observed in the YouTube dataset: most creators stick to a single genre, while a few actively explore different genres (the y-axis is on a log scale). The superiority compared to the baseline is demonstrated in Table~\ref{tab:baseline}.

\subsubsection{\textbf{Creation Activity Alignment}}


Due to the varying creation frequencies of creators in real life, we conduct experiments to evaluate the alignment of CreAgent’s creation activity with real-world creators.
As shown in Figure~\ref{fig:creation_align} (c), we plot the individual-level distribution of creator activity. The deep blue dashed line represents the average creation count per month for 643 creators in the YouTube dataset (in decreasing order), while the light blue area represents the corresponding CreAgent's total creation times under 100 steps' simulation. Since we sample 50 creators per simulation, we conduct 20 experiments to count the average creation times of all 643 creators.
We can observe that the simulation results are very consistent with the actual distribution of creation counts, both exhibiting a long-tail distribution (the y-axis is on a log scale).

\subsubsection{\textbf{Creator Content Alignment}}
\begin{figure}[!ht]
    \centering
\includegraphics[width=0.9\linewidth]{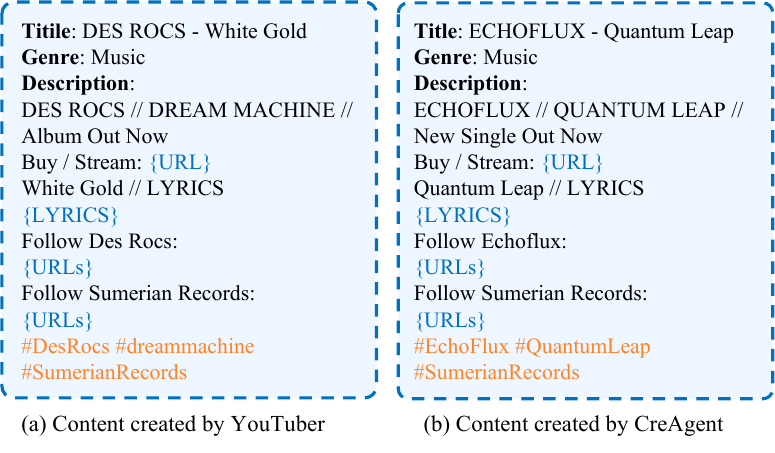}
    \caption{A comparison example between CreAgent and content generated by real-world YouTubers.}
    \label{fig:comparison_content}
\end{figure}

To ensure that our CreAgent can effectively simulate real creators, another crucial aspect is the consistency of the content.
As shown in Figure~\ref{fig:comparison_content}, we compare the content created by a music YouTuber with that generated by the corresponding CreAgent whose profile and creation memory are initialized from the YouTuber. It shows high consistency in form and content, indicating that CreAgent aligns well with real creators in its creation content.

\begin{figure}[ht!]
\centering
\includegraphics[width=1\linewidth]{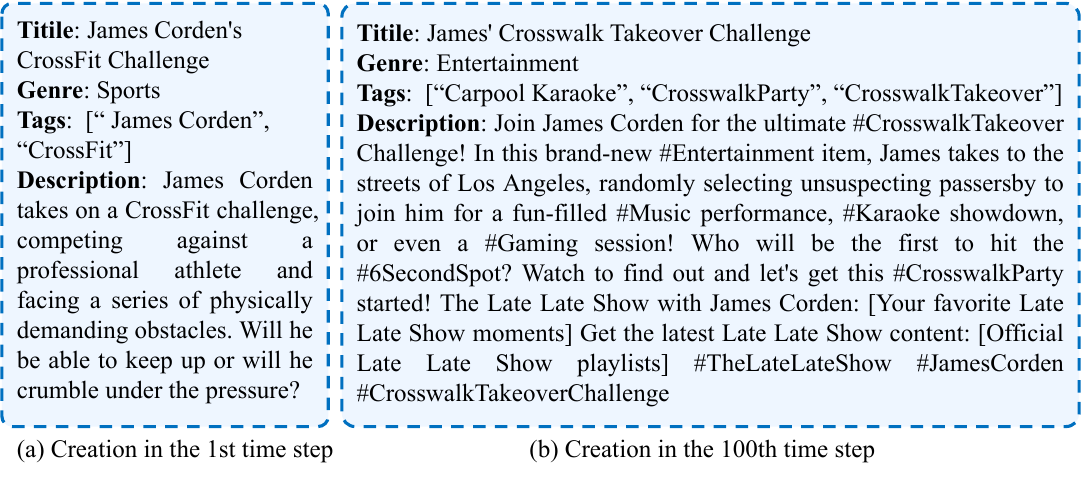}
\caption{An example of the impact of fine-tuning on improving the content generated by CreAgent.}
\label{fig:comparison_content2}
\end{figure}

In real life, creators’ creative skills improve with the increasing number of creations they produce.
To assess the effectiveness of our fine-tuning strategy (see Section~\ref{sec:ppo}) in replicating such a process, we conduct a 100-time-step PPO fine-tuning process and record the creation history of a YouTuber, \textit{James Corden}. 
The creation memory is initialized as empty for the CreAgent to illustrate the creation skill improvement better.
Figure~\ref{fig:comparison_content2} compares the content created by CreAgent at the 1st and 100th step, showing significant improvements in content length, richness, and tag usage.

\subsubsection{\textbf{Ablation Study}}
We conduct an ablation study to investigate the impact of several key modules of CreAgent on creator behavior simulation and alignment (i.e., audience/skill belief, fast-slow thinker), as shown in Table~\ref{tab:baseline}.  We find that both audience belief and skill belief have a significant impact on preference alignment, as they contain information about real creators. Regarding diversity alignment, removing any of these three components greatly affects the results, proving that CreAgent requires all of them to achieve alignment with real-world creative diversity.

\subsection{RQ2: Interaction Behavior Alignment}
\label{sec:strategic_alignment}
Under information asymmetry, the long-term interaction behaviors of CreAgent with the platform (i.e., creation behavior) follow certain patterns, which is supported by renowned behavioral economics theory~\cite{mullainathan2000behavioral_ecnomics}. To evaluate such interaction behavior of CreAgent, we selected the two most fundamental theories from behavioral economics: bounded rational~\cite{selten1990bounded_rationality, conlisk1996boundedrational} and prospect theory~\cite{kahneman2013prospecttheory}.


\subsubsection{\textbf{Bounded Rationality}}

\begin{figure}[!ht]
    \centering
    \includegraphics[width=0.98\linewidth]{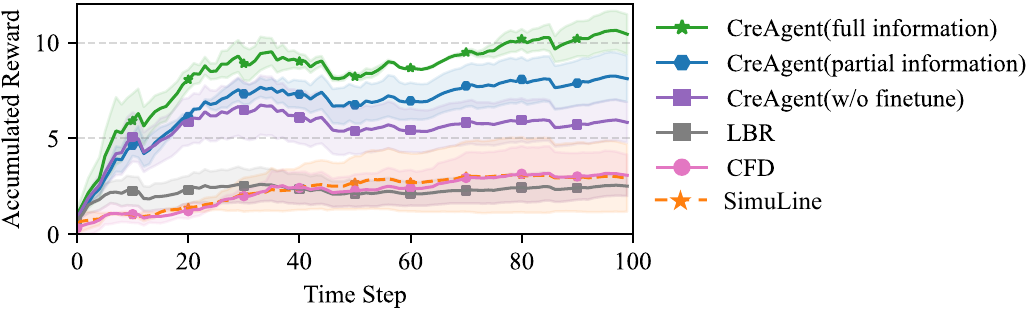}
    \caption{Accumulated reward of creator agents over time steps, normalized by random creation strategy~\cite{shen2023hyperbandit}. } 
    \label{fig:bounded}
\end{figure}
Established behavioral economic studies~\cite{conlisk1996boundedrational, selten1990bounded_rationality} have revealed that \textit{
individuals may not always be fully rational when making decisions. Their choices can be influenced by limited information, leading to suboptimal decisions.}

Figure~\ref{fig:bounded} shows the normalized cumulative rewards~\cite{shen2023hyperbandit} of CreAgent under full and partial information scenarios compared with other baselines.
Full information refers to providing CreAgent with extra information regarding the distribution of user preferences in each genre. We can observe that CreAgent under full information achieves a higher reward than CreAgent under partial information, suggesting that CreAgent under limited information cannot always make fully rational decisions to maximize its reward. Limited information leads CreAgent to make suboptimal choices, which is consistent with human behavior under limited information~\cite{conlisk1996boundedrational}.
Additionally, when comparing CreAgent with the baselines, our approach demonstrates a higher accumulated reward, which indicates that CreAgent leverages human-like analyzing and creation abilities of LLM, exhibiting stronger decision-making and creative capabilities under limited information conditions.



\begin{figure}
    \centering
    \includegraphics[width=1\linewidth]{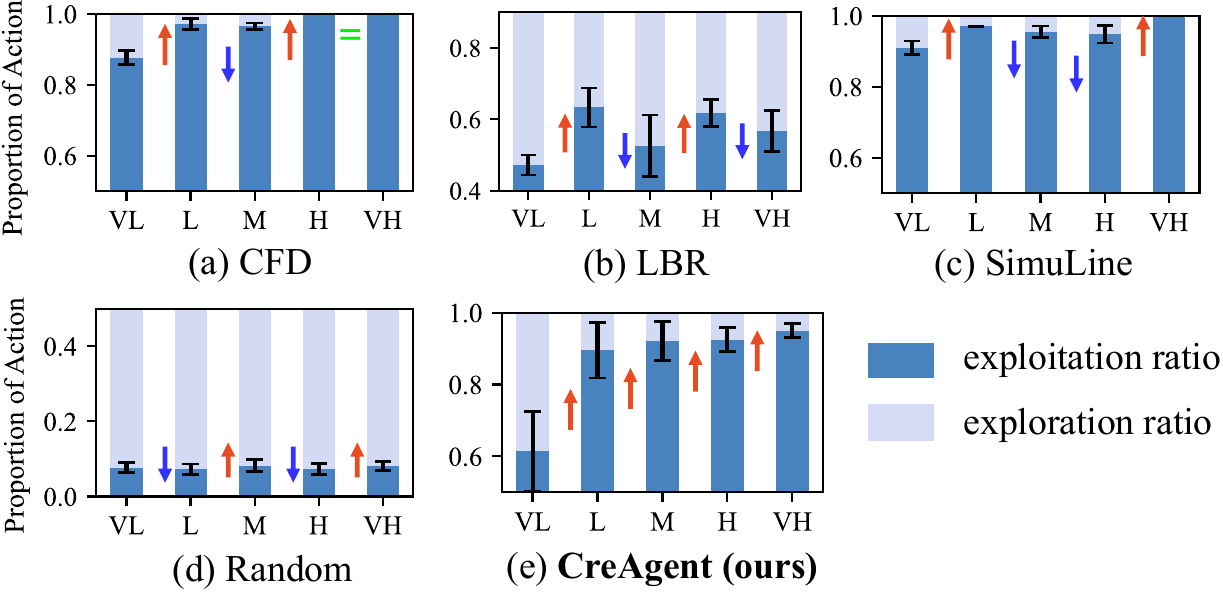}
    \caption{Agent's next action proportion under different reward levels, with VL, L, M, H, and VH for very low, low, medium, high, and very high, respectively. The exploitation ratio of a higher reward level can be higher \textcolor{red}{$\uparrow$}, equal \textcolor{green}{$=$}, or lower \textcolor{blue}{$\downarrow$} to that of a lower reward level.}
    \label{fig:prospect}
\end{figure}

\subsubsection{\textbf{Prospect Theory}}
With only limited information available, the user feedback of  creators' creation action is uncertain and at risk, presenting them with the dilemma of exploration/exploitation. They can either continue creating content they are familiar with (i.e., exploitation) or try unfamiliar fields (i.e., exploration). 
The well-known behavior economic study Prospect Theory~\cite{kahneman2013prospecttheory, kahneman2013prospect_decision} suggests that:
\textit{individuals often exhibit asymmetric behavior in decision-making under uncertainty, where they are more sensitive to losses than to equivalent gains, leading to risk-avoiding in high-reward situations and risk-embracing in low-reward situations.}

 As reported in Figure~\ref{fig:prospect}, we plot the creator agent's next action proportion at different reward levels on the last created item. In Figure~\ref{fig:prospect}(e), when CreAgents under limited information receive a relatively low reward for their previous creation, their creation strategies are more aggressive (i.e., exploration action is nearly 40\%). Conversely, if they receive a high reward, they tend to stick to their existing content strategy to avoid risk, as evidenced by increased exploitation proportion from low to high rewards. 
For comparison, as shown in Figure~\ref{fig:prospect} (a-d), traditional simulation baselines cannot align with behavior economics principles, showing a similar exploration-exploitation ratio under both low-reward and high-reward scenarios.

\subsection{RQ3: Computaional Costs}

\begin{figure}[ht!]
\centering
\includegraphics[width=1\linewidth]{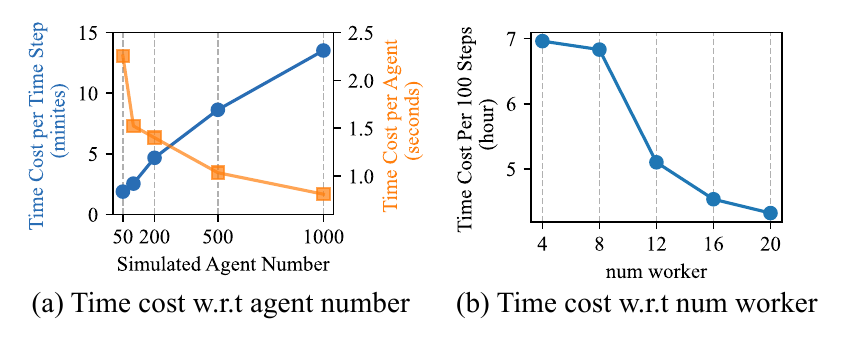}
\caption{ (a) Time cost per agent and time cost per step w.r.t. the number of simulated agents; (b) Time cost for 100 steps' simulation w.r.t number of CPU workers.}
\label{fig:large}
\end{figure}

Though currently limited to small-scale simulations, we discuss computational complexity and costs to show that our simulator can support large-scale simulations with sufficient resources. As shown in Figure~\ref{fig:large} (a), as the number of simulated CreAgents increases, the time cost per step does not rise to an unacceptable level. Even with 1000 agents, the time per step stays under 15 minutes, offering a significant computational cost advantage over online A/B testing, which usually takes weeks or months for long-term evaluations. Moreover, we observe that the average computational cost per agent decreases as the number of agents increases. Additionally, Figure~\ref{fig:large} (b) shows that increasing the number of CPU workers can further reduce the time costs through multi-threaded parallelism, enabling large-scale simulations.


\section{\mbox{Long-term Effect Evaluation}}
\label{sec:RS_evaluation}
In this section, we use CreAgent and the simulation platform to assess the long-term effect of RS on different stakeholders.
Our evaluation focuses on three key long-term objectives that the multi-stakeholder content platform concerns~\cite{ebrahimi2023userengagement, wang2023recagent, xu2023p}: Can different RS models 
(1) improve user engagement (\textbf{RQ4}), (2) protect content creators (\textbf{RQ5}), and (3) enrich content (\textbf{RQ6}) in the long run?

\subsection{Experimental Setups}

\textbf{Evaluation setups.}
\label{sec:simulation_setups}
The experimental setup in this section is identical to Section~\ref{sec:creator_distibution_alignment}.
Note that the first $N_0-1=9$ steps are for environment initialization, followed by 90 evaluation steps starting at $N_0=10$ step.
We evaluate the performance of different ranking models which start recommending at step 1, and re-ranking beginning at step $N_0$.
In this paper, we consider the practical issue of creator retention, assuming a creator leaves if no clicks are received after 5 consecutive creations~\cite{xu2023p, mladenov2020optimizing_long-term}. 
\\
\textbf{Recommendation model selection.}
In this paper, we mainly focus on evaluating three types of recommendation models: (1) basic ranking models (i.e., Random, most popular (Pop), MF~\cite{koren2009mf},  BPR~\cite{rendle2012bpr}, DIN~\cite{zhou2018din}); and (2) provider-fair models (i.e., P-MMF~\cite{xu2023p}, FairRec~\cite{patro2020fairrec}, CPFair~\cite{naghiaei2022cpfair}, TFROM~\cite{wu2021tfrom}, FairCo~\cite{morik2020fairco}); and (3) diversity-aware models (i.e., MMR~\cite{carbonell1998mmr}, APDR~\cite{teo2016apdr}). For MF and BPR, we take user and item ID as inputs. For DIN, we incorporate the embedding of the generated textual content, along with creator ID, genre ID, and the user history interaction sequence, as input features.

\begin{table}[!ht]
    \centering
    \setlength\tabcolsep{5pt} 
        \caption{Rankings model evaluations, with five runs per model using various seeds, display mean and standard deviation with larger and smaller numbers, respectively.}
        \resizebox{1\columnwidth}{!}{
    \begin{tabular}{l|ccc}
    \toprule
    Models  & User Welfare & Creator Retention
    & Content Diversity\\ \midrule
     Random & $6220_{\pm159.8}$ & $\mathbf{0.900}_{\pm0.000}$  & $2.221_{\pm 0.045}$   \\ 
       Pop & $6167_{\pm647.0}$ & $0.500_{\pm0.028}$  & $2.146_{\pm0.095}$  \\ 
       MF & $9322_{\pm 104.7}$ & $0.710_{\pm0.014}$ & $2.190_{\pm0.035}$  \\ 
       BPR & $8554_{\pm353.4}$ & $0.620_{\pm0.000}$  & $\mathbf{2.223}_{\pm0.101}$  \\
       DIN  & $\mathbf{11289}_{\pm 1353}$ & $0.627_{\pm0.012}$  & $1.872_{\pm0.145}$  \\
    \bottomrule
    \end{tabular}
    }
    \label{tab:ranking_model}
\end{table}

\begin{table}[!ht]
    \centering
    \setlength\tabcolsep{5pt} 
        \caption{Evaluation of fairness- and diversity-aware models with DIN as the base model. Setups are same as Table~\ref{tab:ranking_model}.}
          \resizebox{1\columnwidth}{!}{
    \begin{tabular}{l|ccc}
    \toprule
     Models & User Welfare & Creator Retention
    & Content Diversity \\ \midrule
   Base & $11289_{\pm 1353}$ & $0.627_{\pm0.012}$  & $1.872_{\pm0.145}$   \\
           \hline
           \multicolumn{4}{c}{Diversity-aware}  \\  \hline
       +MMR & $11059_{\pm 114.6}$ & $0.68_{\pm 0.028 }$  & $2.017_{\pm 0.008 }$  \\
        +APDR & $13489_{\pm 215.7}$ & $0.720_{\pm 0.057}$  & $1.974_{\pm0.185}$ \\
       \hline
            \multicolumn{4}{c}{Fairness-aware}  \\ \hline
       +FairRec & $14108_{\pm 530.1}$ & $0.840 _{\pm 0.028 }$   & $1.918  _{\pm0.133 }$   \\
       +FairCo & $12749_{\pm 1955}$ & $0.960_{\pm 0.000 }$   & $\mathbf{2.246}_{\pm0.018}$ \\
      +TFROM & $11089_{\pm 326.7}$ & $0.920_{\pm 0.028 }$   &$2.144_{\pm 0.048 }$ \\  
       +P-MMF& $13865_{\pm 225.6}$ & $\mathbf{1.000}_{\pm 0.000}$   & $2.228_{\pm 0.020}$   \\
       +CPFair & $\mathbf{14506}_{\pm 605.3}$
       & $0.940 _{\pm 0.028 }$   & $2.186  _{\pm 0.002 }$ \\

    \bottomrule
    \end{tabular}
}
    \label{tab:reranking_model}
\end{table}

\begin{figure}[!ht]
    \centering
\includegraphics[width=1\linewidth]{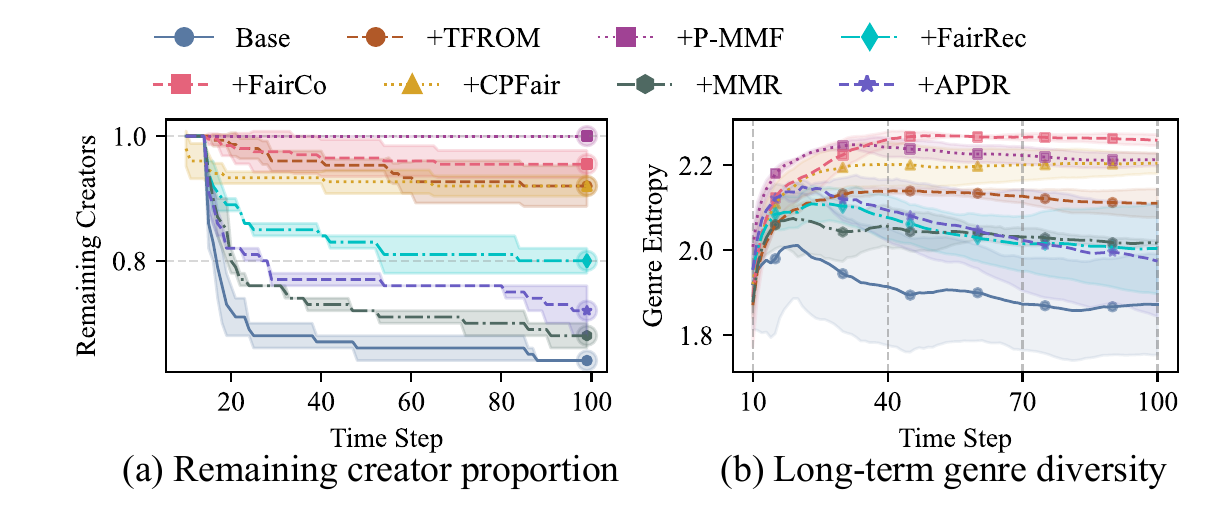}
    \caption{Changes in (a) remaining creators and (b) content enrichment with the increase of time steps under models.}
    \label{fig:remain_creator}
\end{figure}

\subsection{RQ4: User Engagement}
\label{sec:user_engagement}
\textbf{Motivation.}
Long-term user engagement has always been one of the key goals pursued by the platform~\cite{bardhan2022more, ebrahimi2023userengagement}.
We present the Total User Welfare (TUW) metric for evaluation, which is defined as the cumulative user click number:  
\begin{equation}
    \text{Total User Welfare} = \sum_{n=N_0}^{N} \sum_{u\in \mathcal{U}^n} \sum_{i \in \mathcal{I}^n}\mathbf{Y}_{u, i}(n).
\end{equation}
\\
\textbf{Results.}
Table~\ref{tab:ranking_model} shows the long-term user engagement for various ranking models. Simple strategies like Random and Pop perform poorly in this context. While id-based models like MF and BPR offer some improvement, they still fall short in fully grasping user interests. However, feature-based model, such as DIN, more effectively capture long-term user interests, leading to enhanced engagement.

Experiment results in Table~\ref{tab:reranking_model} suggested that diversity and fairness-aware models do not significantly harm long-term user engagement. For instance, MMR only reduced engagement by 2.03\%, and TFROM by 1.77\% compared to the base model. Interestingly, some fairness-aware models (e.g., CPFair), even showed significant improvements, increasing engagement by 28.50\% over the base model.
From the perspective of users, we attribute this phenomenon to the exploration of users’ unknown interests: Short-sighted strategies often focus on users’ existing interests while neglecting their potential interests. Fairness-aware and diversity-aware models will be more effective in discovering users’ potential interests.

\subsection{RQ5: Creator Protection}
\label{sec:creator_protection}
\textbf{Motivation.} 
The protection of creators is crucial for platform growth and user attraction~\cite{bardhan2022more}. However, short-sighted recommendation strategies may prevent niche creators from gaining enough exposure, potentially leading to their departure from the platform~\cite{wu2021tfrom, xu2023p, hohnhold2015focusing}. This can reduce creator diversity and hinder the long-term development~\cite{bardhan2022more}.

To assess the protection of creators of different recommendation models (\textbf{RQ5}), we propose the Creator Retention Rate (CRR) metric, which is defined as the alive creator (i.e., creators who have not yet left) number in the N-th step divided by the alive creator number in the $N_0$-th step.
\\
\textbf{Result.}
Table~\ref{tab:ranking_model} and Table~\ref{tab:reranking_model} show the creator protection performance of different RS models. Aside from the random strategy's high CRR (which gives equal exposure to each item), other strategies have low CRR (below 0.8), suggesting a loss of over 20\% of initial creators after 100 simulation steps.
Figure~\ref{fig:remain_creator} illustrates the remaining creator proportion as the simulation progresses. The base model (DIN) experiences a rapid decline in creators, whereas diversity-aware and fairness-aware algorithms significantly slow this departure. Notably, different fairness algorithms achieve varying degrees of creator protection. For instance, P-MMF~\cite{xu2023p}, designed to boost exposure for worst-exposed creators, retains all creators by the 100th time step, thus effectively safeguarding the platform's creator ecosystem.

\subsection{RQ6: Content Enrichment}
\label{sec:filter_bubble_alleviation}

\textbf{Motivation.}
The enrichment of content is a critical issue for online platforms. Myopic strategies may lead to falling into a filter bubble effect~\cite{nguyen2014explor_filter_bubble, wang2023recagent}, causing decreased user experience, creator exposure unfairness~\cite{wang2023surveyonfairness}, and insufficient exploration of user interests~\cite{chen2021values_of_user_exploration}. 
To measure the level of content enrichment, we follow~\cite{wang2023recagent} and use entropy to quantify the degree of filter bubble effect~\cite{nguyen2014explor_filter_bubble}. We define content genre diversity (CGD) as the average genre entropy received by each user. It can be defined as follows:
\begin{equation}
   \text{Content Diverisity}  = -\frac{1}{\sum_{n=N_0}^N|\mathcal{U}^n|}\sum_{n=n_0}^N \sum_{u\in \mathcal{U}^n} \sum_{g\in\mathcal{G}} p_{u,g} \log \left(p_{u,g} \right),
\end{equation}
where $p_{u,g}=\frac{\sum_{n=N_0}^{N}\sum_{u\in\mathcal{U}^n} \sum_{i \in \mathcal{I}_g}\mathbf{E}_{u,i}(n)}{\sum_{n=N_0}^{N}\sum_{u\in\mathcal{U}^n} \sum_{i}\mathbf{E}_{u,i}(n)}$ is the frequency of genre $g$ recommended to user $u$.
A higher CGD value indicates greater content enrichment and a milder filter bubble effect.
\\
\textbf{Results.}
As reported in Table~\ref{tab:ranking_model}, despite the high long-term user engagement DIN achieves (mentioned in Section~\ref{sec:user_engagement}), it goes through a pronounced filter bubble effect, with a CGD value 15.8\% lower than BPR. We attribute this to its reliance on genre and creator features, which leads to greedy recommendations on the known interests of users.
We also examine the long-term impacts of diversity-aware and fairness-aware models on enriching the content, as illustrated in Figure~\ref{fig:remain_creator} (b) and Table~\ref{tab:reranking_model}.
Compared with the base model whose genre diversity (CGD) declines over time, diversity-aware strategies alleviate this, and fairness-aware strategies notably increase recommendation diversity, as evidenced by FairCo maintaining a consistently high enrichment over time.

\section{Conclusion and Future Works}
In this paper, we propose an LLM-empowered creator simulator to enhance the long-term evaluation of RS under information asymmetry scenarios. Extensive experiments demonstrate CreAgent's effectiveness in aligning with real-world creator behavior and providing valuable insights into the long-term effects of various RS models, including fairness- and diversity-aware algorithms, on the multi-stakeholder environment. 
Although CreAgent offer a promising research direction in multi-stakeholder RS simulation, there are still some limitations. For example, real creator behaviors may involve randomness (not entirely strategic) and multi-modality. We will consider these aspects in the future work.



\appendix

\newpage

\section*{Appendix}

\section{Detail of the Collected Youtube Dataset}
\label{sec:dataset}

Our dataset comprises a comprehensive set of 4,004 content creators, each with a unique influence level, as indicated by their follower counts varying from 10 to 12.32 K. Additionally, we included 1.97 M users and 3.97 M comments in total—spanning 0.19 M distinct items of 14 genres. Due to the inability to collect explicit preference behaviors from specific users (e.g., clicks and likes), we consider user comments as an indication of interest in the item.
Out of privacy protection concerns, we mask specific sensitive information of users and channels.  

Due to resource limitations, the number of creators and users simulated in this article is limited. Therefore, we specifically randomly sampled from the complete YouTube dataset (i.e., Big\_Youtube) to construct a more densely populated Small\_YouTube\ dataset.
To better present the details of the YouTube dataset we collected, we present the statistics of the dataset in Table~\ref{tab:youtube_dataset}.
The dataset has been shared at \textcolor{blue}{\url{https://github.com/shawnye2000/CreAgent}}.

\begin{table}[H]
    \centering
    \caption{Statistics of the collected YouTube Dataset}
    \label{tab:youtube_dataset}
     \begin{tabular}{l|cc}
    \toprule
        Dataset & Big\_Youtube & Small\_Youtube \\ \midrule
        \#Interactions & 3,970,123 & 40,479 \\ 
        \#Users & 1,967,066 & 1,571 \\ 
        \#Items & 186,164 & 64,300 \\ 
        \#Creators & 4,004 & 643 \\ 
         \#Genres & 14 & 14 \\ 
        Inter. Per User & 2.02 & 25.77 \\ 
        Item Per Creator & 46.49 & 100 \\ \bottomrule
    \end{tabular}
\end{table}

\section{Ablation Study on Different LLMs}

We employ two additional LLMs as base models to carry out experiments (that is, Mistral-7B and Qwen2-7B), to verify the capability and effectiveness of CreAgent under these LLMs and explore some new findings.
Due to the time and resource constraints during the rebuttal phase, the number of LLMs we could test is limited. We sincerely apologize for this. In the future, we will explore and evaluate CreAgent with more additional LLMs.

\subsection{Real-world data alignment evaluation}

\begin{table}[!ht]
    \centering
    \caption{Comparison of the divergence between the simulated and real-world distributions using Jensen-Shannon divergence~\cite{menendez1997JS_div}, with genre-level for preference evaluation and individual-level for diversity.}
    \begin{tabular}{l|ll}
    \toprule
       Simulation Method  & Preference  & Diversity \\ \midrule
     Creator Feature Dynamics~\cite{lin2024cfd, zhan2021toward_content_provider_rec} & 0.2537 & 0.7204 \\ 
       Local Better Response~\cite{yao2024uwo} 
       & 0.2833 & 0.6284 \\ 
SimuLine~\cite{zhang2023simuLine} 
& 0.3175 &  0.6949\\ \hline
        CreAgent(Mistral-7B) & 0.2045 & 0.4012 \\ 
        CreAgent(Qwen2-7B) & 0.1917 & 0.3979 \\ 
        CreAgent(LLama3-8B) & 0.1667 & 0.3014 \\ \bottomrule
    \end{tabular}
    
\end{table}

\textbf{Result.}. We observe varying performance when using different LLMs as the base model for our method. In terms of the categories of items created by the simulated creators, the results consistently aligned closely with real-world data. However, in the aspect of creative diversity, some models, such as Mistral-7B, exhibited weaker performance. We hypothesize that this is due to the fact that our prompts were fine-tuned specifically on LLama, leading to potential inconsistencies when applied to other models. Despite this variation, it is notable that while these models may underperform compared to LLama, CreAgent still outperformed the baselines (e.g., CFD[1], LBR[2]).

\subsection{Strategic behavior alignment evaluation}

\begin{table}[!ht]
    \centering
    \begin{tabular}{l|c}
    \toprule
   Simulation Method & Accumulated Reward\\ \midrule
          Creator Feature Dynamics~\cite{lin2024cfd, zhan2021toward_content_provider_rec} & 3.08 \\ 
     Local Better Response~\cite{yao2024uwo} & 2.51 \\ 
     SimuLine~\cite{zhang2023simuLine} 
& 3.04\\ \hline
        CreAgent(Mistral-7B) & 7.32 \\ 
        CreAgent(Qwen2-7B) & 7.98  \\ 
        CreAgent(LLama3-8B) & 8.11 \\ \bottomrule
    \end{tabular}
\end{table}

\textbf{Result.} For the strategic behavior of creator agents, we first conducted experiments to evaluate CreAgent’s reward acquisition capability using different LLMs as base models. All rewards were normalized against the random strategy, following the settings in the paper. As shown in the table, CreAgent consistently demonstrated superior analytical decision-making and creative abilities across almost all base models, achieving higher user rewards. However, we observe that not all base models achieved rewards comparable to LLama3-8B. For instance, Mistral-7B may have limitations in its post-pretraining capabilities, making it less effective at analyzing current user feedback.

\begin{table}[!ht]
    \centering
    \resizebox{1\columnwidth}{!}{
    \begin{tabular}{l|ccccc}
    \toprule
        Simulation Method & Very Low & Low & Medium & High & Very High \\ \midrule
        Random & 0.0431 & 0.0343 & 0.0356 & 0.0547 & 0.052 \\ 
Creator Feature Dynamics~\cite{lin2024cfd, zhan2021toward_content_provider_rec}& 0.8959 & 0.9867 & 0.9746 & 1.0000 & 1.0000 \\
    Local Better Response~\cite{yao2024uwo}& 0.4069 & 0.7009 & 0.6778 & 0.6102 & 0.4583 \\ 
SimuLine~\cite{zhang2023simuLine}& 0.9104&0.9712& 0.9556 & 0.9476&1.0000 \\ \hline
        CreAgent(LLama3-8B) & 0.6138 & 0.8953 & 0.9220 & 0.9250 & 0.9498 \\
        CreAgent(Qwen2-7B) & 0.7272 & 0.8057 & 0.7848 & 0.8170 & 0.8333 \\ 
        CreAgent(Mistral-7B) & 0.4822 & 0.7500 & 0.8471 & 0.8261 & 0.8235 \\ \bottomrule
    \end{tabular}}
\end{table}

\textbf{Result.} We conducted experiments to evaluate the exploration-exploitation balance of creator agents under varying reward levels, assessing whether their behavior aligns with prospect theory. The table highlights the proportion of exploitation actions taken after receiving different levels of user feedback on newly-created items.
 While different LLMs exhibited varying exploration-exploitation levels under different rewards, they all displayed patterns resembling human behavior**. Specifically, agents showed a strong inclination to explore under low rewards and a remarkable tendency to exploit under high rewards, reflecting loss-seeking under low returns and risk aversion under high returns. This behavior sharply contrasts with the random strategy and traditional embedding-based baselines. For instance, CFD shifted to 100\% exploitation at high reward levels, while LBR paradoxically reduced exploitation proportions under high rewards.

We encourage future research to utilize CreAgent and our simulation platform to explore the capabilities and limitations of LLMs in simulating human behavior.

\section{User Alignment Evaluation}
\label{sec:user_alignment}

In this section, we conduct experiments on the user agent employed in our simulation environment to validate how effectively it aligns with real-world user preferences and behaviors.

\subsection{User Item Preference Alignment}
\begin{table}[!ht]
    \centering
    \caption{User Item Preference Alignment}
    \begin{tabular}{c|cccc}
    \toprule
        1:m & Accuracy  & Precision  & Recall  & F1 Score  \\ \midrule
        1:1 & $0.630_{\pm0.031}$  & $\mathbf{0.658 }_{\pm0.023 }$  & $0.603 _{\pm0.070 }$  & $\mathbf{0.589}_{\pm0.054 }$ \\ 
        1:2 & $0.598_{\pm0.037 }$  & $0.461 _{\pm0.042 }$  & $0.523 _{\pm0.059 }$  &$0.460 _{\pm 0.047  }$    \\
        1:3 & $0.622 _{\pm0.015 }$  &  $0.373 _{\pm 0.022 }$   & $0.520 _{\pm0.031 }$ & $0.404 _{\pm0.020 }$    \\ 
        1:9 & $\mathbf{0.653} _{\pm0.040 }$  & $0.276  _{\pm0.036 }$    & $\textbf{0.740}_{\pm0.063 }$  & $0.358_{\pm0.039 }$  \\ \bottomrule
    \end{tabular}
    \label{tab:user_item_pref}
\end{table}
\textbf{Motivation.} To validate how well generative agents align with the real-world preferences, we utilze the user agents to differentiate between items that actual users have engaged with and those they have not.
Specifically, a total of 200 agents will each be randomly assigned 20 items. Among these, the ratio between items the user has interacted with (i.e., $\mathbf{Y}_{u,i}(0) = 1$) but was not utilized for profile initialization and items the user has not interacted with (i.e., $\mathbf{Y}_{u,i}(0) = 0$) is set as 1:m, with $m \in \{1, 2, 3, 9\}$. Under this setting, user agent responses (i.e., $\mathbf{Y}_{u,i}(n) = 1, n\in [1,2, \cdots, N]$) to recommended items are considered binary discrimination, taking values between 0 and 1. Then, we compute the accuracy, precision, recall, and f1-score metric to show its performance.

\textbf{Results.} Table~\ref{tab:user_item_pref} reports the empirical discrimination results across various metrics. The best performance for each metric is highlighted in bold and marked with an asterisk. 
We observe that:
The generative user agents consistently identify items that align well with user preferences, maintaining around 65.3\% accuracy and 74\% recall even when faced with 18 (i.e., $1:m=1:9$) distracting items. This high performance is attributed to the personalized profiles that accurately reflect users' true interests, demonstrating the agents' ability to encapsulate real preferences and highlighting the viability of LLM-powered generative agents in recommendation systems.

In our item-by-item recommendation setting, the user agents do not tend to click on a certain number of items in the recommendation list, as mentioned in~\cite{zhang2024agent4rec}. However, our user agent can ensure a high level of Recall (above 0.5) and accuracy (nearly above 0.6) when 1:m decreases.
This indicates that our user agent, however, maintains a certain tendency towards identifying positive examples even when the proportion of similar items decreases, which may result in some negative items being identified as positive.
We attribute this failure to LLM's inherent hallucinations that agents tend to consistently pick a set number of items. However, we emphasize that in the subsequent simulation results with recommendation algorithms, the recommendation list length is set to 5, hence a substantial proportion of recommended items align with user preferences, thereby endorsing high trustworthiness in those simulation outcomes.

\subsection{User Genre Preference Alignment}
\begin{figure}[H]
    \centering
\includegraphics[width=1\linewidth]{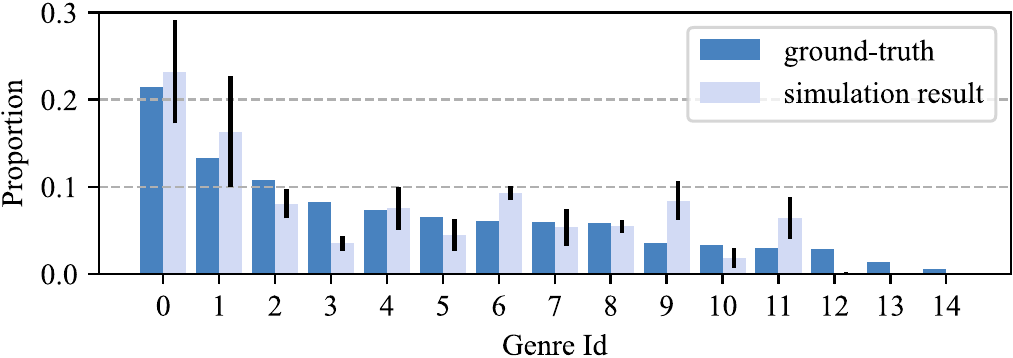}
    \caption{Comparison between the distributions of ground-truth and agent-simulated genre preference.}
    \label{fig:user-interest-alignment}
\end{figure}

\textbf{Motivation.} In a real-world RS, users have unique interests in different genres of items. Since these interests drive user actions such as viewing, clicking, and liking, it is crucial for our user agents to align with the preferences of real-world users.
Specifically, we aim to align the interest distribution of our user agents with the distribution observed in the real-world dataset we have collected.
\\
\textbf{Results.}
To verify the consistency between the interest distribution of our user agent and that of real-world users, we conducted experiments using our CreAgent framework. We initialized user and creator profiles with the YouTube dataset we collected, utilizing DIN as the recommendation model and simulating 100 time steps. Figure~\ref{fig:user-interest-alignment} (a) shows the interest distribution of users from the real-world dataset, where users who clicked more than five times on a category are considered to favor that category. The x-axis represents item genres, sorted from highest to lowest proportionally. Figure~\ref{fig:user-interest-alignment} (b) illustrates the interest distribution of our user agent, where we compute the proportion of clicks for each genre.
From the comparison, we can see that our user agent ultimately achieved an interest distribution similar to that of real-world users. However, we are unable to perfectly replicate the relative differences across certain genres. For example, in the \textit{Howto \& Style} (H\&S) genre, the user agent exhibited a higher preference than real users, while extremely low preference is observed in \textit{Sports} (s). 
We attribute this to the LLM's extensive prior knowledge of genres, which causes the agent to exhibit a stronger preference for certain genres of items.

\section{Details of the Prompts}
\label{app:prompt}

\subsection{Prompts for Profile Summarization}
\label{app:profile_prompts}
\begin{tcolorbox}[colback=gray!5, colframe=gray!60, coltitle=white, breakable,fonttitle=\bfseries, title=Designed Prompt for Social Identity Summarization]

\textbf{Prompt}: You are a content creator on \{platform name\} and your name is \{creator name\}. Here is the basic information about the content you have previously created.

Recent created content: {title:<title>, genre:<genre>, description:<description>}

Created content genre (the genres you have created in the past and their respective proportions): \{created genre proportion\} 

Creation frequency (the average number of items you create each day): \{creation time per day\}

Please summarize your social identity in the following format: [Social Identity]: <the specific identity>. For example, [Social Identity]: movie enthusiast.
\end{tcolorbox}

\begin{tcolorbox}[colback=gray!5, colframe=gray!60, coltitle=white, breakable,fonttitle=\bfseries, title=Designed Prompt for Intrinsic Motivation Summarization]

\textbf{Prompt}: You are a content creator on \{platform name\} and your name is \{creator name\}. Here is the basic information about the content you have previously created.

Follower number: {follower number}

Average views per video: {average views}.

Recent created content: {title:<title>, genre:<genre>, description:<description>}

Recent interaction with users (your recent interaction records with the audience in the comments section.): \{recent comments\} 

Creation frequency (the average number of items you create each day): \{creation time per day\}

Intrinsic motivation refers to whether your purpose for creating content is for profit or simply for sharing. Please summarize your intrinsic motivation in the following format: [Intrinsic Motivation]: <the specific motivation>. For example, [Intrinsic Motivation]: profit. 
\end{tcolorbox}

\subsection{Prompts for Creation Module}
\label{app:creation_prompts}

\begin{tcolorbox}[colback=gray!5, colframe=gray!60, coltitle=white, breakable,fonttitle=\bfseries, title=Designed Prompt for Fast Thinker]

\textbf{Prompt}: You are a content creator on YouTube and your nickname is \{name\}. 

\{profile text of $P_c^m$,and $P_c^a$\}

Based on the analysis: \{$\textbf{A}^{\text{exp}}_c$\}, please create ONE new content for \{name\} that fits user's interest.

You can refer to the creation history of \{name\}: \{$f(\mathcal{M}^{\text{cre}}_c)$\}

Response in JSON dictionary format.
Write {{"name": [item name], "genre": genre1|genre2|....,  "tags": [tag1, tag2, tag3], "description": "item description text"}})
\end{tcolorbox}

\begin{tcolorbox}[colback=gray!5, colframe=gray!60, coltitle=white, breakable,fonttitle=\bfseries, title=Designed Prompt for Slow Thinker]

\textbf{Prompt}: You are a content creator on YouTube and your nickname is \{name\}. 

\{profile text of $P_c^m$,and $P_c^a$\}

The average utility per item of each genre \{name\} has created is as below: \{$\mathbf{B^{\text{aud}}_c}$\}. ([unknown] means the item genre \{name\} have not explored.

Recently, \{name\} created an item of genre \{$g_i$\}, and receives \{$\mathbf{z}_i(n)$\} utility.

Due to the statistical data, \{name\}'s profile and \{name\}'s 
familiarity on each genre: {$\mathbf{B^{\text{skill}}_c}$}, \{name\} must choose one of the two actions below to obtain more user clicks:

(1) [EXPLORE] Create content in a new genre that has not been explored before, which means other genres may have a larger audience and more opportunities to profit. But it might not be \{name\}'s area of expertise and requires greater effort to create.

(2) [EXPLOIT] Sticking to creating content of a familiar genre, which means \{name\} will leverage his creative expertise to build a stable brand identity. But it might limit \{name\}'s audience reach and lead to insufficient income.

To explore a new genre, write: [EXPLORE]:: <genre name>. If so, give the specific genre name chosen from {unknown cates}.

To stick to familiar genres, write: [EXPLOIT]:: <genre name>. If so, give the specific genre name chosen from {known cates}.
                          
  Let's think step by step. Please answer concisely and strictly follow the output rules.

\textbf{Responses Example of $\mathbf{A}^{\text{exp}}_c$}: [EXPLORE]: Entertainment
\end{tcolorbox}


\newpage
\bibliographystyle{ACM-Reference-Format}
\bibliography{sample-base}


\end{document}